\documentclass[manuscript=article]{achemso}

\setkeys{acs}{usetitle=true}

\usepackage{subfigure}
\usepackage{amsmath}
\usepackage{xcolor}
\usepackage[version=3]{mhchem}
\usepackage{siunitx}
\title[Graph Networks as a Universal Machine Learning Framework for Molecules and Crystals]{Graph Networks as a Universal Machine Learning Framework for Molecules and Crystals}
\author{Chi Chen}
\affiliation[UCSD]{Department of NanoEngineering, University of California San Diego, 9500 Gilman Dr, Mail Code 0448, La Jolla, CA 92093-0448, United States}
\author{Weike Ye}
\affiliation[UCSD]{Department of NanoEngineering, University of California San Diego, 9500 Gilman Dr, Mail Code 0448, La Jolla, CA 92093-0448, United States}
\author{Yunxing Zuo}
\affiliation[UCSD]{Department of NanoEngineering, University of California San Diego, 9500 Gilman Dr, Mail Code 0448, La Jolla, CA 92093-0448, United States}
\author{Chen Zheng}
\affiliation[UCSD]{Department of NanoEngineering, University of California San Diego, 9500 Gilman Dr, Mail Code 0448, La Jolla, CA 92093-0448, United States}
\author{Shyue Ping Ong}
\email{ongsp@eng.ucsd.edu}
\affiliation[UCSD]{Department of NanoEngineering, University of California San Diego, 9500 Gilman Dr, Mail Code 0448, La Jolla, CA 92093-0448, United States}

\date{\today}

\begin{document}

\begin{abstract}
Graph networks are a new machine learning (ML) paradigm that supports both relational reasoning and combinatorial generalization. Here, we develop universal MatErials Graph Network (MEGNet) models for accurate property prediction in both molecules and crystals. We demonstrate that the MEGNet models outperform prior ML models such as the SchNet in 11 out of 13 properties of the QM9 molecule data set. Similarly, we show that MEGNet models trained on $\sim 60,000$ crystals in the Materials Project substantially outperform prior ML models in the prediction of the formation energies, band gaps and elastic moduli of crystals, achieving better than DFT accuracy over a much larger data set.  \textcolor{black}{We present two new strategies to address data limitations common in materials science and chemistry. First, we demonstrate a physically-intuitive approach to unify four separate molecular MEGNet models for the internal energy at 0 K and room temperature, enthalpy and Gibbs free energy into a single free energy MEGNet model by incorporating the temperature, pressure and entropy as global state inputs. Second, we show that the learned element embeddings in MEGNet models encode periodic chemical trends and can be transfer-learned from a property model trained on a larger data set (formation energies) to improve property models with smaller amounts of data (band gaps and elastic moduli).}
\end{abstract}

\maketitle

\section{Introduction}

Machine learning (ML)\cite{michalski2013machine, LeCun2015a} has emerged as a powerful new tool in materials science,\cite{Mueller2016, Ramprasad2017, Pilania2013, Ward2016, Rupp2012a, Hautier2010, Xie2017, Schutt2017, Bartok2010, Butler2018, Ye2018a, bartok2017machine} driven in part by the advent of large materials data sets from high-throughput electronic structure calculations\cite{Jain2013, saal2013materials, curtarolo2012aflow, nomad2018} and/or combinatorial experiments\cite{chan2015combinatorial, xiang2014high}. Among its many applications, the development of fast, surrogate ML models for property prediction has arguably received the most interest for its potential in accelerating materials design\cite{mansouri2018machine, oliynyk2017discovery} as well as accessing larger length/time scales at near-quantum accuracy.\cite{Behler2007a,Bartok2010,deringer2018data, Thompson2015, Wood2017a, artrith2017efficient, Chen2017a}

The key input to any ML model is a description of the material, which must satisfy the necessary rotational, translational and permutational invariances as well as uniqueness. For molecules, graph-based representations\cite{Bonchev1991} are a natural choice. \textcolor{black}{This graph representation concept was then successfully applied  to predict  molecular properties.\cite{duvenaud2015convolutional, coley2017convolutional}  } Recently, \citet{Faber2017} have benchmarked different features in combination with models extensively on the QM9 data set.\cite{Ramakrishnan2014} They showed that the graph-based deep learning models\cite{Kearnes2016a, li2015gated} generally outperform classical ML models with various features. Furthermore, graph-based models are generally less sensitive to the choice of atomic descriptors, unlike traditional feature engineering-based ML models. For example, Sch{\"u}tt et al.\cite{Schutt2017,Schutt2018} achieved state-of-the-art performance on molecules using only the atomic number and atom coordinates in a graph-based neural network model. \citet{gilmer2017neural} later proposed the message passing neural network (MPNN) framework that includes the existing graph models with differences only in their update functions. 

Unlike molecules, descriptions of crystals must account for lattice periodicity and additional space group symmetries. In the crystal graph convolutional neural networks (CGCNN) proposed by \citet{Xie2017}, each crystal is represented by a crystal graph, and invariance with respect to permutation of atomic indices and unit cell choice are achieved through convolution and pooling layers. They demonstrated excellent prediction performance on a broad array of properties, including formation energy, band gap, Fermi energy and elastic properties.

Despite these successes, current ML models still suffer from several limitations. First, it is evident that most ML models have been developed on either molecular or crystal datasets. A few notable exceptions are the recently reported SchNet\cite{Schutt2018} and an update of the MPNN \cite{jorgensen2018neural} which have been tested on both molecules and crystals, although in both cases performance evaluation on crystals is limited to formation energies only. Second, current models lack a description of global state (e.g., temperature), which are necessary for predicting state-dependent properties such as the free energy. Last but not least, data availability remain a critical bottleneck for training high-performing models for some properties. For example, while there are $\sim 69,000$ computed formation energies in the Materials Project, \cite{Jain2013} there are only $\sim 6,000$ computed elastic constants.

In this work, we aim to address all these limitations. We propose graph networks\cite{Battaglia2018} with global state attributes as a general, composable framework for \textcolor{black}{quantitative structure-state-property relationship} prediction in materials, i.e., both molecules and crystals. \textcolor{black}{Graph networks can be shown to be a generalization/superset of previous graph-based models such as the CGCNN and MPNN; however, because graph networks are not constrained to be neural network-based, they are different from the afore-mentioned models.} We will demonstrate that our MatErials Graph Network (MEGNet) models outperform prior ML models in the prediction of multiple properties on the $\sim 131,000$ molecules in the QM9 data set\cite{Ramakrishnan2014} and $\sim$ 69,000 crystals in the Materials Project.\cite{Jain2013} \textcolor{black}{We also present a new physically-intuitive strategy to unify multiple free energy MEGNet models into a single MEGNet model by incorporating state variables such as temperature, pressure and entropy as global state inputs, which provides for multi-fold increase in the training data size with minimal increase in number of model parameters. Finally, we demonstrate how interpretable chemical trends can be extracted from elemental embeddings trained on a large data set, and these elemental embeddings can be used in transfer learning to improve the performance of models with smaller data quantities.}

\section{Methods}

\subsubsection{MEGNet Formalism}

Graph networks were recently proposed by \citet{Battaglia2018} as a general, modular framework for ML that supports both relational reasoning and combinatorial generalization. Indeed, graph networks can be viewed as a superset of the previous graph-based neural networks, though the use of neural networks as function approximators is not a prerequisite. Here, we will outline the implementation of MEGNet models for molecules and crystals, with appropriate modifications for the two different material classes explicitly described. Throughout this work, the term ``materials'' will be used generically to encompass molecules to crystals, while the more precise terms ``molecules'' and ``crystals'' will be used to refer to collections of atoms without and with lattice periodicity, respectively.  

\begin{figure}[htp]
\includegraphics[width=1 \textwidth]{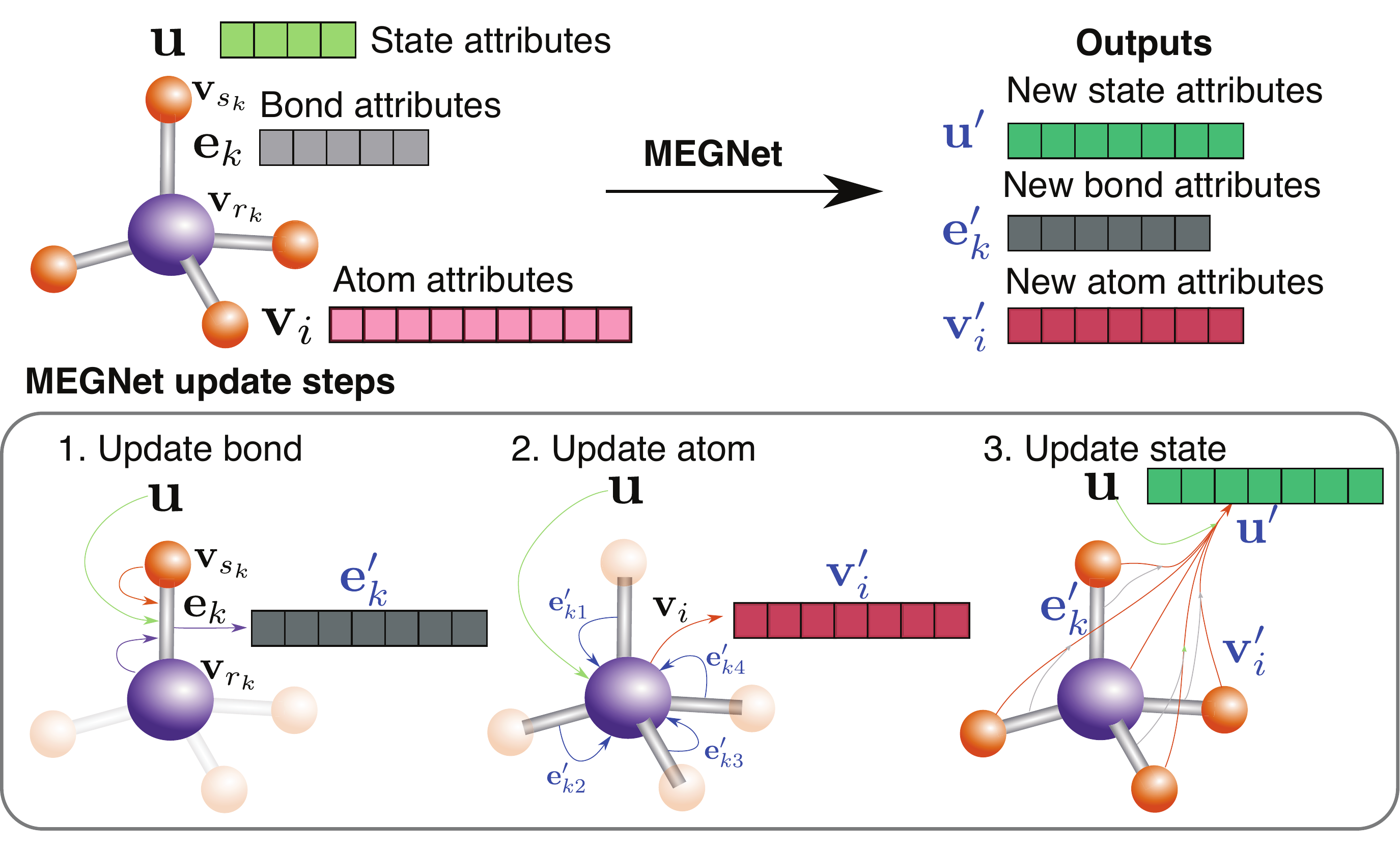}
\caption{\label{fig:gnmodule} Overview of a MEGNet module. The initial graph is represented by the set of atomic attributes $V = \{\mathbf{v_i}\}_{i=1:N^v}$, bond attributes $E = \{(\mathbf{e_k}, r_k, s_k)\}_{k=1:N^e}$ and global state attributes $\mathbf{u}$. In the first update step, the bond attributes are updated. Information flows from atoms that form the bond, the state attributes and the previous bond attribute to the new bond attributes. Similarly, the second and third steps update the atomic and global state attributes, respectively, by information flow among all three attributes. The final result is a new graph representation.} 
\end{figure}

Let $V$, $E$ and $\mathbf{u}$ denote the atomic (node/vertex), bond (edge) and global state attributes respectively. For molecules, bond information (e.g., bond existence, bond order, etc.) is typically provided as part of the input definition. For crystals, a bond is loosely defined between atoms with distance less than certain cut-off. Following the notation of \citet{Battaglia2018}, $V$ is a set of $\mathbf{v_i}$, which is an atomic attribute vector for atom $i$ in a system of $N^v$ atoms. $E = \{(\mathbf{e_k}, r_k, s_k)\}_{k=1:N^e}$ are the bonds, where $\mathbf{e_k}$ is the bond attribute vector for bond $k$, $r_k$ and $s_k$ are the atom indices forming bond $k$, and $N^e$ is the total number of bonds. Finally, $\mathbf{u}$ is a global state vector storing the molecule/crystal-level or state attributes (e.g., the temperature of the system).

A graph network module (Figure \ref{fig:gnmodule}) contains a series of update operations that map an input graph $G=(E, V, \mathbf{u})$ to an output graph $G'=(E', V', \mathbf{u}')$. First, the attributes of each bond $(\mathbf{e_k}, r_k, s_k)$ are updated using attributes from itself, its connecting atoms (with indices $r_k$ and  $s_k$) and the global state vector $\mathbf{u}$, as follows:

\begin{equation}
  \mathbf{e}'_k =  \phi_e\left(\mathbf{v}_{s_k} \bigoplus \mathbf{v}_{r_k} \bigoplus \mathbf{e}_k \bigoplus  \mathbf{u}\right)\label{eqn:bond_update}
\end{equation}
where $\phi_e$ is the bond update function and $\bigoplus$ is the concatenation operator. Next, the attributes of each atom $\mathbf{v_i}$ are updated using attributes from itself, the bonds connecting to it, and the global state vector $\mathbf{u}$, as follows:

\begin{eqnarray}
  \mathbf{\bar v}^e_i =  \frac{1}{N^e_i}\sum_{k=1}^{N^e_i} \{\mathbf{e}'_k\}_{r_k = i}\label{eqn:aggregation}\\
  \mathbf{v}'_i = \phi_v\left( \mathbf{\bar v}^e_i \bigoplus \mathbf{v}_i \bigoplus \mathbf{u} \right)\label{eqn:atomupdate2}
\end{eqnarray}
where $N_i^e$ is the number of bonds connected to atom $i$, and $\phi_v$ is the atom update function. The aggregation step (Equation \ref{eqn:aggregation}) acts as a local pooling operation that takes the average of bonds that connect to the atom $i$. 

The first two update steps contain localized convolution operations that rely on the atom-bond connectivity. One can imagine that if more graph network modules are stacked, atoms and bonds will be able to ``see'' longer distances, and hence, longer range interactions can be incorporated even if the initial distance cut-off is small to reduce the computational task. 

Finally, the global state attributes $\mathbf{u}$ are updated using information from itself and all atoms and bonds, as follows:

\begin{eqnarray}
\label{eqn:stateupdate}
  \mathbf{\bar u}^e =  \frac{1}{N^e}\sum_{k=1}^{N^e} \{\mathbf{e}'_k\}  \\\label{eqn:bond_to_state}
    \mathbf{\bar u}^v =  \frac{1}{N^v}\sum_{i=1}^{N^v} \{\mathbf{v}'_i\}  \\\label{eqn:atom_to_state}
  \mathbf{u}' = \phi_u\left( \mathbf{\bar u}^e \bigoplus \mathbf{\bar u}^v \bigoplus \mathbf{u} \right) \label{eqn:state_update}
\end{eqnarray}
where $\phi_u$ is the global state update function. In addition to providing a portal to input state attributes (e.g., temperature), $\mathbf{u}$ also acts as the global information placeholder for information exchange on larger scales. 

The choice of the update functions $\phi_e$, $\phi_v$ and $\phi_u$ largely determines the model performance in real tasks. In this work, we choose the $\phi$s to be multi-layer perceptrons with two hidden layers (Equation \ref{eqn:multi_perceptron}), given their ability to be universal approximators for non-linear functions.\cite{Hornik1989} 

\begin{equation}
  \phi(\mathbf{x}) = \mathbf{W_3} (\zeta (\mathbf{W_2}(\zeta (\mathbf{W_1}\mathbf{x} + \mathbf{b_1}) ) + \mathbf{b_2})) +  \mathbf{b_3} \label{eqn:multi_perceptron}
\end{equation}
where $\zeta$ is the modified softplus function\cite{Schutt2017} acting as nonlinear activator, $\mathbf{W}$s are the kernel weights and $\mathbf{b}$s are the biases. Note that the weights for atom, bond and state updates are different. Each fully-connected layer will be referred as a ``dense'' layer using keras\cite{chollet2015keras} terminology. 

\begin{figure}[htp]
\includegraphics[width=1 \textwidth]{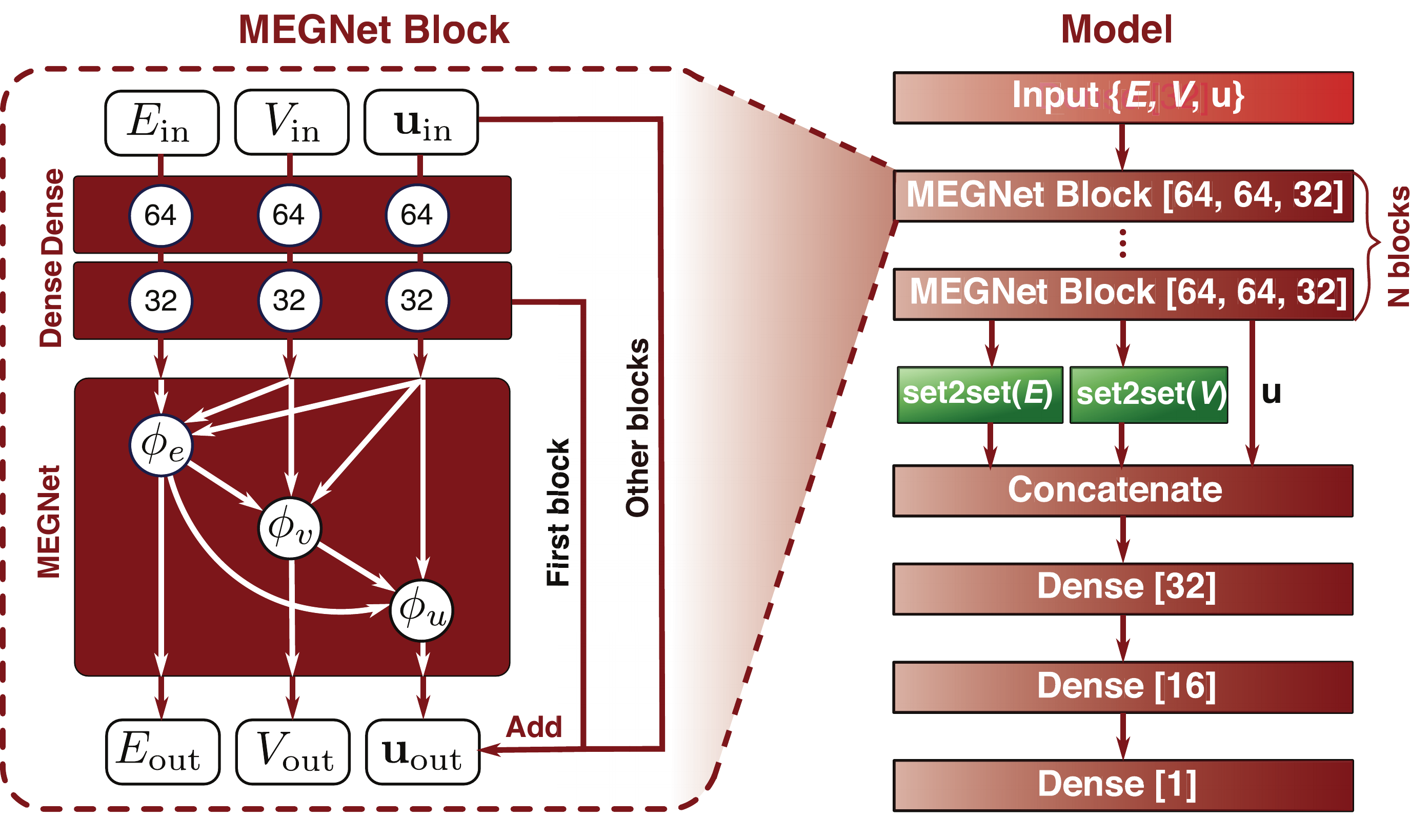}
\caption{\label{fig:model_arch}Architecture for the MEGNet model. Each model is formed by stacking MEGNet blocks. In the readout stage, a \textit{set2set} neural network is used to reduce sets of atomic and bond vectors into a single vector. The numbers in brackets are the number of hidden neural units for each layer. Each MEGNet block contains a MEGNet layer as well as two dense layers. The ``Add'' arrows are skip-connections to enable deep model training.} 
\end{figure}

To increase model flexibility, two dense layers are added before each MEGNet module to pre-process the input. This approach has been found to increase model accuracy. We define the combination of the two dense layers with a MEGNet module as a MEGNet block, as shown in Figure \ref{fig:model_arch}. The block also contains residual net-like\cite{he2016deep} skip connections to enable deeper model training and reduce over-fitting. Multiple MEGNet blocks can stacked to make more expressive models.  In the final step, a readout operation reduces the output graph to a scalar or vector. In this work, the order-invariant \textit{set2set} model\cite{Vinyals2015} that embeds a set of vectors into one vector is applied on both atomic and bond attributes sets. After the readout, the atomic, bond and state vectors are concatenated and passed through multi-layer perceptrons to generate the final output. The overall model architecture is shown in Figure \ref{fig:model_arch}. If the atom features are only the integer atomic numbers, an embedding layer is added after the atom inputs $V$.

\subsection{Atomic, Bond and State Attributes}

Table \ref{tbl:feature} summarizes the full set of atomic, bond and state attributes used as inputs to the MEGNet models. The molecule attributes are similar to the ones used in the benchmarking work by \citet{Faber2017}. For crystals, only the atomic number and spatial distance are used as atomic and bond attributes, respectively.

\begin{table}
\caption{\label{tbl:feature} Atomic, bond and state attributes used in the graph network models.}
\begin{tabular}{lllp{0.5\linewidth}}
\hline
System &Level& Attributes name & Description \\
\hline
Molecule & Atom & Atom type & H, C, O, N, F (one-hot).\\
& & Chirality & R or S (one-hot or null).\\
& & Ring sizes & For each ring size (3-8), the number of rings that include this atom. If atom is not in a ring, this field is null. \\
& & Hybridization & $sp$, $sp^2$, $sp^3$ (one-hot or null). \\
& & Acceptor & Whether the atom is an electron acceptor (binary) \\
& & Donor & Whether the atom donates electrons (binary)
\\
& & Aromatic & Whether the atom belongs to an aromatic ring. (binary) \\

&Bond & Bond type & Single, double, triple, or aromatic (one-hot or null). \\
& &Same ring & Whether the atoms in the bond are in the same ring (binary).\\
& & Graph distance & Shortest graph distance between atoms (1-7). \textcolor{black}{This is a topological distance. For example, a value of 1 means that the two atoms are nearest neighbors, while a value of 2 means they are second nearest neighbors etc.} \\
& & Expanded distance & \textcolor{black}{Distance $r$ valued on Gaussian basis $\exp(-(r-r_0)^2/\sigma^2)$ where $r_0$ takes values at 20 locations linearly placed between 0 and 4, and the width $\sigma = 0.5$.} \\

& State & Average atomic weight & Molecular weight divided by number of atoms (float). \\
& & Bonds per atom & Average number of bonds per atom (float).\\
\hline
Crystal & Atom & Z & The atomic number of element (1-94)\\
& Bond & Spatial distance & Expanded distance with Gaussian basis $\exp(-(r-r_0)^2/\sigma^2)$ centered at 100 points linearly placed between 0 and 5 and $\sigma = 0.5$.\\
& State & Two zeros & Placeholder for global information exchange. \\
\hline
\end{tabular}

\end{table}

\subsection{Data Collections}

The molecule data set used in this work is the QM9 data set\cite{Ramakrishnan2014} processed by \citet{Faber2017} It contains the B3LYP/6-31G(2df,p)-level DFT calculation results on 130,462 small organic molecules containing up to 9 heavy atoms.

The crystal data set comprises the DFT-computed energies and band gaps of 69,640 crystals from the Materials Project\cite{Jain2013} obtained via the Python Materials Genomics (\textit{pymatgen})\cite{Ong2013a} interface to the Materials Application Programming Interface (API)\cite{ong2015materials} on June 1, 2018. We will designate this as the MP-crystals-2018.6.1 data set to facilitate future benchmarking and comparisons as data in the Materials Project is constantly being updated. The crystal graphs were constructed using a radius cut-off of 4 \AA. Using this cut-off, 69,239 crystals do not form isolated atoms and are used in the models. All crystals were used for the formation energy model and the metal against non-metals classifier, while a subset of 45,901 crystals with finite band gap was used for the band gap regression. A subset of 5830 structures have elasticity data that do not have calculation warnings and will be used for elasticity models. 

\subsection{Model Construction and Training}

A customized Python version of MEGNet was developed using the \textit{keras} API\cite{chollet2015keras} with the \textit{tensorflow} backend.\cite{Abadi2016} Since molecules and crystals do not have the same number of atoms, we assemble batches of molecules/crystals into a single graph with multiple targets to enable batch training. The Adam optimizer\cite{Kingma2014} was used with an initial learning rate of 0.001, which is reduced to 0.0001 during later epochs for tighter convergence. 

Each data set is divided into three parts - training, validation and test. For the molecule models, 90\% of the data set was used for training and the remaining were divided equally between validation and test. For the crystal formation energy models, 60,000 crystals were used for training and the remaining were divided equally between validation and test for direct comparison to the work of \citet{Schutt2018}. For the band gap classification models and elastic moduli models, an 80:10:10 split was applied. All models were trained on the training set, and the configuration and hyperparameters with the lowest validation error were selected. Finally, the test error is calculated. \textcolor{black}{During training, the validation error is monitored and the training is stopped when the validation error does not improve for 500 consecutive epochs. The models were trained on Nvidia GTX 1080Ti GPUs. On average, it takes 80 and 110 seconds per epoch for each molecular and crystal model, respectively. Most models reach convergence within 1000 epochs. However, models for $U_0$, $U$, $H$, $G$, and $\left<R^2\right>$ require 2000 to 4000 epochs.} In crystals, the embedding dimension is set to 16. The elemental embeddings trained on the formation energy using one MEGNet block was transferred to the band gap regression model and kept fixed. We use the same architecture featuring three MEGNet blocks in the models for crystals.  

\subsection{Data and Model Availability}

To ensure reproducibility of the results, the MP-crystals-2018.6.1 data set used in this work have been made available as a JavaScript Object Notation file at \url{https://figshare.com/articles/Graphs_of_materials_project/7451351}. The graph network modules and overall models have also been released as open-source code in a Github repository at \url{https://github.com/materialsvirtuallab/megnet}.

\section{Results}

\subsection{Performance on QM9 Molecules}

\begin{table}
  \caption{\label{tbl:qm9} Comparison of mean absolute errors (MAEs) of 13 properties in the QM9 data set for different models. The ``Benchmark'' column refers to the best model in the work by \citet{Faber2017}, and the ``Target'' column refers to the widely-accepted thresholds for ``chemical accuracy''.\cite{Faber2017} \textcolor{black}{The standard deviations in the MAEs for the MEGNet-Full models over three randomized training:validation:test splits are also provided.}}
  \footnotesize
  \begin{tabular}{llllllll}
    \hline
    Property  & Units & MEGNet-Full\textsuperscript{*} & MEGNet-Simple\textsuperscript{**} &Schnet\cite{Schutt2018}  & enn-s2s\cite{gilmer2017neural} & Benchmark \cite{Faber2017} & Target\\
    & & (This Work) & (This Work) & & & & \\
    \hline
    $\boldsymbol{\epsilon}_{\mathrm{HOMO}}$ & $\mathrm{eV}$ & \textbf{0.038$\pm$0.001} & 0.043 &0.041  & 0.043 & 0.055 \textsuperscript{\emph{a}} & 0.043\\
   $\boldsymbol{\epsilon}_{\mathrm{LUMO}}$ & $\mathrm{eV}$ & \textbf{0.031$\pm$0.000}& 0.044&0.034 & 0.037 & 0.064 \textsuperscript{\emph{a}} & 0.043 \\
   $\boldsymbol{\Delta \epsilon}$ & $\mathrm{eV}$ & \textbf{0.061$\pm$0.001} & 0.066&0.063 & 0.069 & 0.087 \textsuperscript{\emph{a}} & 0.043 \\
    $\mathrm{ZPVE}$ &$\mathrm{meV}$ & \textbf{1.40$\pm$0.06} & 1.43&1.7& 1.5 & 1.9 \textsuperscript{\emph{c}} & 1.2\\
    $\mu$ &D  & 0.040$\pm$0.001& 0.050&0.033& \textbf{0.030} & 0.101 \textsuperscript{\emph{a}} & 0.1\\
    $\alpha$ & $\mathrm{bohr}^3$ & \textbf{0.083$\pm$0.001} & 0.081& 0.235 & 0.092 & 0.161 \textsuperscript{\emph{b}} & 0.1\\
    $\left<R^2\right>$ & $\mathrm{bohr}^2$ & 0.265$\pm$0.001& 0.302& \textbf{0.073} & 0.180 & - & 1.2\\
    $U_0$ & $\mathrm{eV}$ & \textbf{0.009$\pm$0.000} & 0.012& 0.014 & 0.019 & 0.025 \textsuperscript{\emph{c}} & 0.043\\
    $U$ & $\mathrm{eV}$ &\textbf{0.010$\pm$0.000} & 0.013&0.019 & 0.019 & - & 0.043\\ 
    $H$ & $\mathrm{eV}$ & \textbf{0.010$\pm$0.000} & 0.012&0.014 & 0.017 & - & 0.043\\
    $G$ & $\mathrm{eV}$ & \textbf{0.010$\pm$0.000} & 0.012&0.014 & 0.019 & - & 0.043\\
    $C_v$ & $\mathrm{cal(mol K)^{-1}}$ & \textbf{0.030$\pm$0.001} & 0.029& 0.033 & 0.040 & 0.044 \textsuperscript{\emph{c}} & 0.05\\
    $\omega_1$ & $\mathrm{cm^{-1}}$ & \textbf{1.10$\pm$0.08} & 1.18&- & 1.9 & 2.71\textsuperscript{\emph{d}} & 10 \\
    \hline
  \end{tabular}
  \begin{flushleft}
  $\boldsymbol{\epsilon}_{\mathrm{HOMO}}$: highest occupied molecular orbital; $\boldsymbol{\epsilon}_{\mathrm{LUMO}}$: lowest unoccupied molecular orbital; $\boldsymbol{\Delta \epsilon}$: energy gap; $\mathrm{ZPVE}$: zero point vibrational energy; $\mu$: dipole moment; $\alpha$: isotropic polarizability; $\left<R^2\right>$: electronic spatial extent; $U_0$: internal energy at 0 K; $U$: internal energy at 298 K; $H$: enthalpy at 298 K; $G$: Gibbs free energy at 298 K; $C_v$: heat capacity at 298 K; $\omega_1$: highest vibrational frequency.\\ 
   \textsuperscript{\emph{*}} Full MEGNet models using all listed features in Table \ref{tbl:feature}. The optimized models for $\mathrm{ZPVE}$, $\left<R^2\right>$, $\mu$ and $\omega_1$ contain five, five, three and one MEGNet blocks, respectively, while the optimized models for all other properties uses two MEGNet blocks.\\
   \textsuperscript{\emph{**}} Simple MEGNet models using only the atomic number as atomic feature, expanded distance as bond features and no dummy state features. All models contain three MEGNet blocks. \\ 
  \textsuperscript{\emph{a}} Graph convolution with molecular graph feature\cite{Kearnes2016a}.\\
  \textsuperscript{\emph{b}} Gated-graph neural network with molecular graph feature\cite{li2015gated}. \\
  \textsuperscript{\emph{c}} Kernel-ridge regression with histogram of distance, angles and dihedrals (HDAD) features.\\
  \textsuperscript{\emph{d}} Random forest model with bonds angles machine learning (BAML) feature.
  \end{flushleft}
  
\end{table}

Table \ref{tbl:qm9} compares the mean absolute errors (MAEs) of 13 properties for the different models and the convergence plots with number of training data are in Figure S1. It can be seen that the MEGNet models using the full set of attributes (``Full'' column in Table \ref{tbl:qm9}) outperforms the state-of-art SchNet\cite{Schutt2018} and MPNN enn-s2s models\cite{gilmer2017neural} in all but two of the properties - the norm of dipole moment $\mu$ and the electronic spatial extent $R^2$. Out of the 13 properties, only the errors on zero-point energy (ZPVE) (1.40 meV) and band gap ($\boldsymbol{\Delta \epsilon}$) (0.060 eV) exceed the thresholds for chemical accuracy. The errors of various properties follow Gaussian distributions, as shown in Figure S2.

We note that the atomic and bond attributes in Table \ref{tbl:feature} encode redundant information. For example, the bond type can usually be inferred from the bonding atoms and the spatial distance. We therefore developed ``simple'' MEGNet models that utilize only the atomic number and spatial distance as the atomic and bond attributes, respectively. \textcolor{black}{These are the same attributes used in the crystal MEGNet models.} From Table \ref{tbl:qm9}, we may observe that these simple MEGNet models achieve largely similar performance as the full models, with only slightly higher MAEs that are within chemical accuracy and still outperforming prior state-of-the-art models in 8 of the 13 target properties. It should be noted, however, that the convergence of the ``simple'' models are slower than the ``full'' models for certain properties (e.g., $\mu$, $ZVPE$). This may be due to the models having to learn more complex relationships between the inputs and the target properties. 

\subsection{Unified Molecule Free Energy Model}

To achieve the results presented in Table \ref{tbl:qm9}, one MEGNet model was developed for each target, similar to previous works.\cite{Schutt2018,gilmer2017neural} However, this approach is extremely inefficient when multiple targets are related by a physical relationship and should share similar features. For instance, the internal energy at 0K ($U_0$) and room temperature ($U$), enthalpy ($H = U + PV$) and Gibbs free energy ($G = U + PV - TS$) are all energy quantities that are related to each other by temperature ($T$), pressure ($P$), volume ($V$) and entropy ($S$). To illustrate this concept, we have developed a combined free energy model for $U_0$, $U$, $H$ and $G$ for the QM9 data set by incorporating the temperature, pressure (binary) and entropy (binary) as additional global state attributes in $\mathbf{u}$, i.e., (0, 0, 0), (298, 0, 0), (298, 1, 0) and (298, 1, 1) for $U_0$, $U$, $H$ and $G$, respectively. Using the same architecture, this combined free energy model achieves an overall MAE of 0.010 eV for the four targets, which is comparable to the results obtained using the separate MEGNet models for each target.

In principle, the combined free energy  model should be able to predict free energies at any temperature given sufficient training data. Indeed, the predicted $U$ at 100 K and 200 K match well with our DFT calculations (see Figure S3), even though these data points were not included in the training data. However, the predicted $H$ and $G$ at the same temperatures show large deviations from the DFT results. We hypothesize that this is due to the fact that only one temperature data for these quantities exist in the training data and that the addition of $H$ and $G$ data at multiple temperatures into the training data would improve the performance of the unified free energy MEGNet model.

\subsection{Performance on Materials Project Crystals}

\begin{table}
  \caption{ \label{tbl:MEGNetperformancemp}Comparison of the MAEs in the formation energy $E_f$, band gap $E_g$, bulk modulus $K_{VRH}$, shear modulus $G_{VRH}$ and and metal/non-metal classification between MEGNet models and prior works on the Materials Project data set. The number of structures in the training data is in parentheses. \textcolor{black}{The standard deviations in the MAEs for the MEGNet models over three randomized training:validation:test splits are also provided.}}
  \begin{tabular}{lllll}
    \hline
      & Units & MEGNet &SchNet\cite{Schutt2018}  & CGCNN\cite{Xie2017}\\
    \hline
 Elements & & 89 & 89 & 87\\
 $E_f$ & eV atom$^{-1}$ & \textbf{0.028$\pm$0.000} (60000) & 0.035 (60000) & 0.039 (28046) \\
 $E_g$ & eV & \textbf{0.33$\pm$0.01} (36720) & - & 0.388 (16485) \\
 $K_{VRH}$ & $\log_{10}$ (GPa) & \textbf{0.050$\pm$0.002} (4664)& - & 0.054 (2041)\\
 $G_{VRH}$ & $\log_{10}$ (GPa) & \textbf{0.079$\pm$0.003} (4664) & - & 0.087 (2041)\\
 Metal classifier & - & 78.9\%$\pm$1.2\% (55391) & - & 80\% (28046)\\
 Non-metal classifier & - & 90.6\%$\pm$ 0.7\% (55391)& - & \textbf{95\%} (28046)\\
 \hline
  \end{tabular}
\end{table}

\begin{figure}[htp]
\includegraphics[width=1 \textwidth]{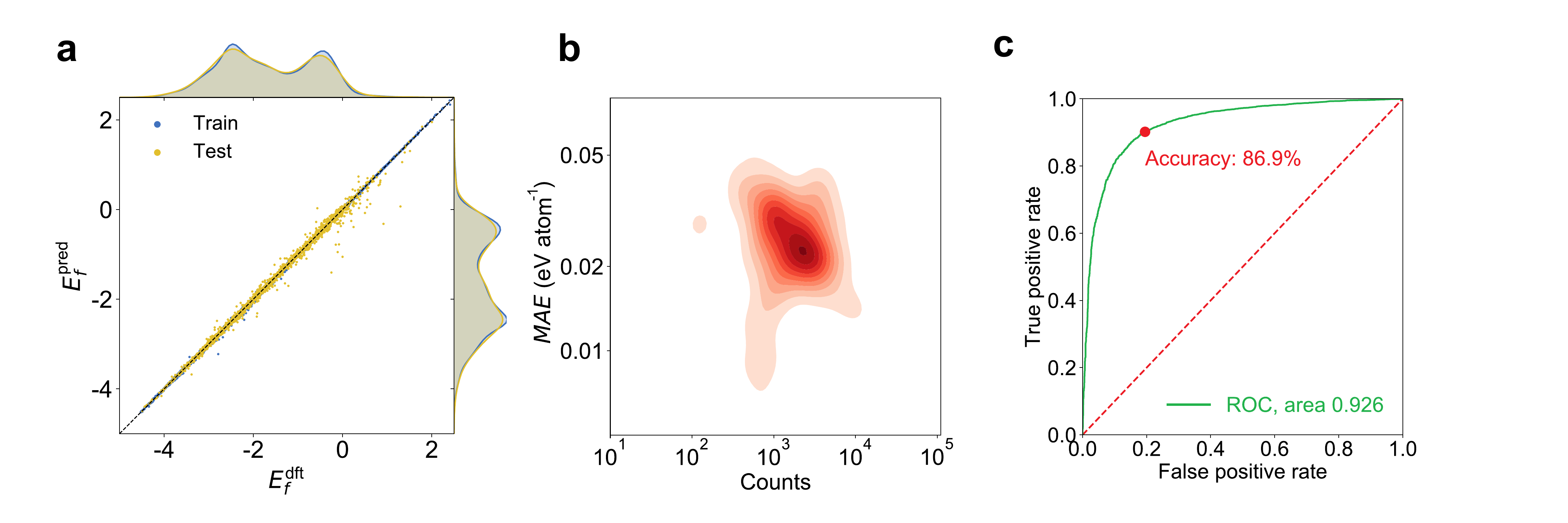}
\caption{\label{fig:formation}Performance of MEGNet models on the Materials Project data set. (a) Parity plots for the formation energy of the training and test data sets. (b) Plot of average MAE for each element against number of training structures containing that element. (c) Receiver operating characteristics (ROC) curve for test data for MEGNet classifier trained to distinguish metals against non-metals. } 
\end{figure}

Table \ref{tbl:MEGNetperformancemp} compares the performance of the MEGNet models against the SchNet\cite{Schutt2018} and CGCNN models\cite{Xie2017}. \textcolor{black}{The convergence of formation energy model is shown in Figure S4.} We may observe that the MEGNet models outperform both the SchNet and CGCNN models in the MAEs of the formation energies $E_f$, band gap $E_g$, bulk modulus $K_{VRH}$ and shear modulus $G_{VRH}$. It should be noted that these results - especially the prediction of $E_g$ and the metal/non-metal classifiers - are achieved over much diverse datasets than previous works, and the prediction error in $E_f$, $E_g$, $K_{VRH}$ and $G_{VRH}$ are well within the DFT errors in these quantities\cite{kirklin2015open,lany2008semiconductor, jain2011high, crowley2016resolution, de2015charting}. The MEGNet models, similar to the SchNet models, utilize only one atomic attribute (atomic number) and one bond attribute (spatial distance), while nine attributes were used in the CGCNN model. We also found that transferring the elemental embeddings from the $E_f$ model, which was trained on the largest data set, significantly accelerates the training and improves the performance of the $E_g$, $K_{VRH}$ and $G_{VRH}$ models. For example, an independently-trained model (without transfer learning) for $E_g$ has a higher MAE of 0.38 eV. 

\textcolor{black}{We note that the data set used in the development of the CGCNN model is significantly smaller than that of MEGNet or SchNet, despite all three models having obtained their data from the Materials Project. The reason is that crystals with warning tags or without band structures were excluded from the CGCNN model training. Using this exclusion strategy and a similar training data size, the MEGNet models for formation energy and band gap have MAEs of 0.032 eV atom$^{-1}$ and 0.35 eV, respectively. The accuracies for metal and non-metal classifiers are increased to 82.7\% and 93.1\% respectively.}

\textcolor{black}{There are also non-graph-based crystal ML models such as the JARVIS-ML model\cite{choudhary2018machine} and the AFLOW-ML model\cite{isayev2017universal}. The MAEs of the JARVIS-ML models\cite{choudhary2018machine} for formation energy, band gap, bulk moduli and shear moduli are 0.12 eV atom$^{-1}$, 0.32 eV, 10.5 GPa and 9.5 GPa, respectively, while the MAEs of AFLOW-ML models\cite{isayev2017universal} for band gap, bulk moduli and shear moduli are 0.35 eV, 8.68 GPa and 10.62 GPa, respectively. However, these ML models are developed with very different data sets (for example, the JARVIS-DFT database contains formation energies, elastic constants and band gaps for bulk and 2D materials computed using different functionals), and are therefore not directly comparable to the MEGNet, SchNet or CGCNN models, which are all trained using Materials Project data.}

Figures \ref{fig:formation}a and b provide a detailed analysis of the MEGNet model performance on $E_f$. The parity plot (Figure \ref{fig:formation}a) shows that the training and test data are similarly well-distributed, and consistent model performance is achieved across the entire range of $E_f$. We have performed a sensitivity analysis of our MEGNet $E_f$ model to various hyperparameters. Increasing the radius cut-off to 6 \AA\, slightly increases the MAE to 0.03 eV atom$^{-1}$. Using one or five MEGNet blocks instead of three result in MAEs of 0.033 and 0.027 eV atom$^{-1}$, respectively. Hence, we can conclude that our chosen radius cut-off of 4 \AA\,  and model architecture comprising three MEGNet blocks are reasonably well-optimized. Figure \ref{fig:formation}b plots the average test MAEs for each element against the number of training structure containing that element. In general, the greater the number of training structures, the lower the MAE for structures containing that element.  Figure \ref{fig:formation}c shows the receiver operating characteristic (ROC) curve for the metal/non-metal classifier. The overall test accuracy is 86.9\%, and the area under curve for the receiver operation conditions is 0.926.

\section{Discussion}

It is our belief that the separation of materials into molecules and crystals is largely arbitrary, and a true test of any structured representation is its ability to achieve equally good performance in property prediction in both domains. We have demonstrated that graph networks, which provide a natural framework for representing the attributes of atoms and the bonds between them, are universal building blocks for highly accurate prediction models. Our MEGNet models, built on graph network concepts, show significantly improved accuracies over prior models in most properties for both molecules and crystals. 

A key advance in this work is the demonstration of the incorporation of global state variables to build unified models for related properties. A proof of concept is shown in our unified molecule free energy MEGNet model, which can successfully predict the internal energy at multiple temperatures, enthalpy and Gibbs free energy with temperature, entropy and pressure as global state variables. \textcolor{black}{This stands in sharp contrast to the prevailing approach in the materials ML community of building single-purpose models for each quantity, even if they are related to each other by well-known thermodynamic relationships. The unification of related models has significant advantages in that one can achieve multi-fold increases in training data with minimal increase in model complexity, which is particularly important given the relatively small datasets available in materials science.}

\subsection{Interpretability}

\begin{figure}[htp]
\includegraphics[width=1 \textwidth]{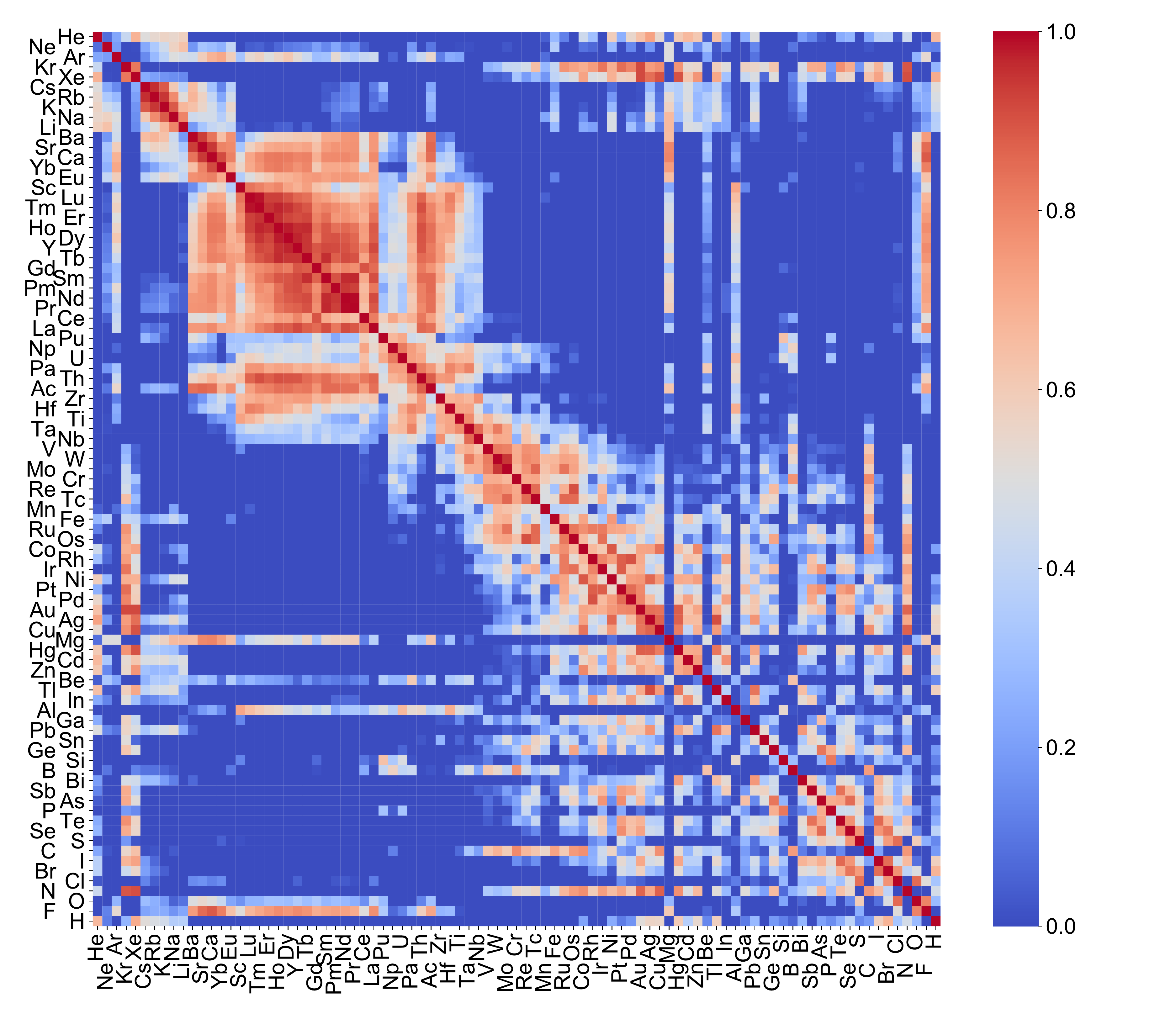}
\caption{\label{fig:embedding}Pearson correlations between elemental embedding vectors. Elements are arranged in order of increasing Mendeleev number\cite{pettifor1988structure} for easier visualization of trends.} 
\end{figure}

For chemistry and materials science applications, a particularly desirable feature for any representation is interpretability and reproduction of known chemistry intuition\textcolor{black}{\cite{zhou2018learning}. To this end, we have extracted the elemental embeddings from the MEGNet model for crystal formation energy. As shown in Figure \ref{fig:embedding}, the correlations between the elemental embeddings correctly recover the trends in the periodic table of the elements. For example, the alkaline, alkali, chalcogen, halogen, lanthanoid, transition metals, post transition metals, metalloid and actinoid show highest similarities within their groups. It is important to note that the extracted trends reproduce well-known ``exceptions'' in the periodic arrangement of atoms as well. For example, the fact that Eu and Yb do not follow the lanthanoids but are closer to alkaline earth elements (Figure S6) is in good agreement with chemical intuition and matches well with the structure graphs proposed by Pettifor.\cite{pettifor1988structure} Furthermore, these trends are obtained from the diverse Materials Project dataset encompassing most known crystal prototypes and 89 elements, rather than being limited to specific crystal systems\cite{xie2018hierarchical,willatt2018data}}.

Such embeddings obtained from formation energy models are particularly useful for the development of models to predict stable new compounds. \citet{Hautier2010a} previously developed an ionic substitution prediction algorithm using data mining, which has been used successfully in the discovery of several new materials\cite{hautier2011novel, wang2018mining}. The ion similarity metric therein is purely based on the presence of ions in a given structural prototype, a somewhat coarse-grained description. Here, the MEGNet models implicitly incorporate the local environment of the site and should in principle better describe the elemental properties and bonding relationships. We note that with more MEGNet blocks, the contrast of the embeddings between atoms are weaker, as shown in Figure S5. The two-dimensional t-SNE plots\cite{Maaten2008} confirm these conclusions, as shown in Figure S6. This is because with more blocks, the environment seen by the atom spans a larger spatial region, and the impact of geometry becomes stronger, which obscures the chemical embeddings.

\subsection{Composability}

A further advantage of the graph network based approach is its modular and composable nature. In our MEGNet architecture, a single block captures the interactions between each atom and its immediate local environment (defined via specified bonds in the molecule models and a radius cutoff in the crystal models). Stacking multiple blocks allows for information flow, and hence, capturing of interactions, across larger spatial distances.

We can see this effect in the MEGNet models for the QM9 data set, where different number of blocks are required to obtain good accuracy for different properties. For most properties, two blocks are sufficient to achieve MAEs within chemical accuracy. However, more blocks are necessary for the zero-point vibrational energy (five), electronic spatial extent (five) and dipole moment (three), which suggests that it is important to capture longer-ranged interactions for these properties. In essence, the choice of number of MEGNet blocks for a particular property model boils down to a consideration of the range of interactions necessary for accurate prediction, or simply increasingly the number of blocks until convergence in accuracy is observed.

\subsection{Data Limitations and Transfer Learning}

The critical bottleneck in building graph networks models, like all other ML models, is data availability. For instance, we believe the inability of the unified free energy MEGNet model to accurately predict $H$ and $G$ at 100 K and 200 K is largely due to the lack of training data at those temperatures.  Similarly, a general inverse relationship can be seen between the number of training structures and the average MAE in formation energies of the crystals in Figure \ref{fig:formation}b. 

Besides adding more data (which is constrained by computational cost as well as chemistry considerations), another avenue for improvement is to use ensemble models. We tested this hypothesis by training two independent three-block MEGNet models and used the average as the ensemble prediction for the formation energies of the Materials Project data set. The MAE reduces from 0.028 eV atom$^{-1}$ for single MEGNet model to 0.024 eV atom$^{-1}$ for the ensemble MEGNet model.

\textcolor{black}{Yet another approach to address data limitations is transfer learning\cite{s2018outsmarting, altae2017low}}, and we have demonstrated an instructive example of how this can be applied in the case of the crystal MEGNet models. Data quantity and quality is a practical problem for many materials properties. Using the Materials Project as an example, the formation energy data set comprises $\sim 69,000$ crystals, i.e., almost all computed crystals in the database. However, only about half of these have non-zero band gaps. Less than 10\% crystals in Materials Project have computed elastic constants, due to the high computational effort in obtaining these properties. By transferring the elemental embeddings, which encode the learned chemical trends from the much larger formation energy data set, we were able to efficiently train the band gap and elastic moduli MEGNet models and achieve significantly better performance than prior ML models. We believe this to be a particularly effective approach that can be extended to other materials properties with limited data availability.

\section{Conclusion}

To conclude, we have developed materials graph network models that are universally high performing across a broad variety of target properties for both  molecules and crystals. Graphs are a natural choice of representation for atoms and the bonds between them, and the sequential update scheme of graph networks provide a natural approach for information flow among atoms, bonds and global state. Furthermore, we demonstrate two advances - incorporation of global state inputs and transfer learning of elemental embeddings - in this work that extend these models further to state-dependent and data-limited properties. These generalizations address several crucial limitations in the application of ML in chemistry and materials science, and provide a robust foundation for the development of general property models for accelerating materials discovery.  

\begin{acknowledgement}
This work is supported by the Samsung Advanced Institute of Technology (SAIT)'s Global Research Outreach (GRO) Program. The authors also acknowledge data and software resources provided by the Materials Project, funded by the U.S. Department of Energy, Office of Science, Office of Basic Energy Sciences, Materials Sciences and Engineering Division under Contract No. DE-AC02-05-CH11231: Materials Project program KC23MP, and computational resources provided by Triton Shared Computing Cluster (TSCC) at the University of California, San Diego, the National Energy Research Scientiﬁc Computing Centre (NERSC), and the Extreme Science and Engineering Discovery Environment (XSEDE) supported by National Science Foundation under Grant No. ACI-1053575.
\end{acknowledgement}

\begin{suppinfo}
MEGNet error distributions on the QM9 dataset;  Energy predictions at different temperatures; 
Elemental embeddings for one and five-MEGNet-block models; t-SNE visualization of elemental embeddings for one, three and five-MEGNet-block models.
\end{suppinfo}

\bibliography{refs}

\end{document}